\documentclass[fleqn,usenatbib,useAMS]{mnras}
\usepackage[utf8]{inputenc} % Set encoding to UTF-8

\usepackage[T1]{fontenc}
\usepackage{ae,aecompl}

\usepackage{graphicx} % Including figure files
\usepackage{amsmath}  % Advanced maths commands
\usepackage{bm}       % For bold math symbols

\makeatletter
\def\tabular{\def\@halignto{}%
 \def\hline{\noalign{\ifnum0=`}\fi
  \vskip 3pt
  \hrule \@height \arrayrulewidth
  \vskip 3pt
  \futurelet \@tempa\@xhline}%
 \def\fullhline{\noalign{\ifnum0=`}\fi
  \vskip 3pt
  \hrule \@height \arrayrulewidth
  \vskip 3pt
  \futurelet \@tempa\@xhline}%
 \def\@xhline{\ifx\@tempa\hline
   \vskip -6pt
   \vskip \doublerulesep
  \fi
  \ifnum0=`{\fi}}%
  \def\@arrayrule{\@addtopreamble{\hskip -.5\arrayrulewidth
                                  \hskip .5\arrayrulewidth}}%
\@tabular
}
\makeatother

\title[The atmosphere of WASP-98\,b]{An optical transmission spectrum of the transiting hot Jupiter in the metal-poor WASP-98 planetary system}
\author[L. Mancini et al.]
{L. Mancini$^{1,2}$\thanks{E-mail: mancini@mpia.de},
M. Giordano$^{3,4}$, %\thanks{E-mail: mancini.giordano@le.infn.it},
P. Molli\`{e}re$^{1}$,
J. Southworth$^{5}$,
R. Brahm$^{6,7}$,
\newauthor
S. Ciceri$^{8}$,
Th. Henning$^{1}$ \\
$^{1}$Max Planck Institute for Astronomy, K\"{o}nigstuhl 17, 69117 -- Heidelberg, Germany \\
$^{2}$INAF -- Osservatorio Astrofisico di Torino, via Osservatorio 20, 10025 -- Pino Torinese, Italy \\
$^{3}$Dipartimento di Matematica e Fisica `\emph{E. De Giorgi}', Universit\`{a} del Salento, Via per Arnesano, CP 193, I-73100 Lecce, Italy \\
$^{4}$INFN, Sezione di Lecce, Via per Arnesano, CP 193, I-73100 Lecce, Italy \\
$^{5}$Astrophysics Group, Keele University, Staffordshire, ST5 5BG, UK \\
$^{6}$Instituto de Astrof\'{i}sica, Pontificia Universidad Cat\'{o}lica de Chile, Av. Vicu\~{n}a Mackenna 4860, 7820436 Macul, Santiago, Chile \\
$^{7}$Millennium Institute of Astrophysics, Av. Vicu\~{n}a Mackenna 4860, 7820436 Macul, Santiago, Chile \\
$^{8}$Department of Astronomy, Stockholm University, SE-106 91 Stockholm, Sweden
}

\date{Accepted XXX. Received YYY; in original form ZZZ}

\pubyear{2015}

\begin{document}
\label{firstpage}
\pagerange{\pageref{firstpage}--\pageref{lastpage}}
\maketitle

\begin{abstract}
The WASP-98 planetary system represents a rare case of a hot Jupiter hosted by a metal-poor main-sequence star. We present a follow-up study of this system based on multi-band photometry and high-resolution spectroscopy. Two new transit events of WASP-98\,b were simultaneously observed in four passbands ($g^{\prime}, r^{\prime}, i^{\prime}, z^{\prime}$), using the telescope-defocussing technique, yielding eight high-precision light curves with point-to-point scatters of less than $1$\,mmag. We also collected three spectra of the parent star with a high-resolution spectrograph, which we used to remeasure its spectral characteristics, in particular its metallicity. We found this to be very low, ${\rm [Fe/H]} =-0.49 \pm 0.10$, but larger than was previously reported, ${\rm [Fe/H]} =-0.60 \pm 0.19$. We used these new photometric and spectroscopic data to refine the orbital and physical properties of this planetary system, finding that the stellar and planetary mass measurements are significantly larger than those in the discovery paper. In addition, the multi-band light curves were used to construct an optical transmission spectrum of WASP-98\,b and probe the characteristics of its atmosphere at the terminator. We measured a lower radius at $z^{\prime}$ compared with the other three passbands. The maximum variation is between the $r^{\prime}$ and $z^{\prime}$ bands, has a confidence level of roughly $6\sigma$ and equates to 5.5 pressure scale heights. We compared this spectrum to theoretical models, investigating several possible types of atmospheres, including hazy, cloudy, cloud-free, and clear atmospheres with titanium and vanadium oxide opacities. We could not find a good fit to the observations, except in the extreme case of a clear atmosphere with TiO and VO opacities, in which the condensation of Ti and V was suppressed. As this case is unrealistic, our results suggest the presence of an additional optical-absorbing species in the atmosphere of WASP-98\,b, of unknown chemical nature.
\end{abstract}


%%%%%%%%%%%%%%%%%%%%%%%%%%%%%%%%%%%%%%%%%%%%%%%%%%

%%%%%%%%%%%%%%%%% BODY OF PAPER %%%%%%%%%%%%%%%%%%

%%%%%%%%%%%%%%%%%%%%%%%%%%%
\section{Introduction}
\label{sec:introduction}
\begin{figure}
  \centering
  \includegraphics[width=\columnwidth]{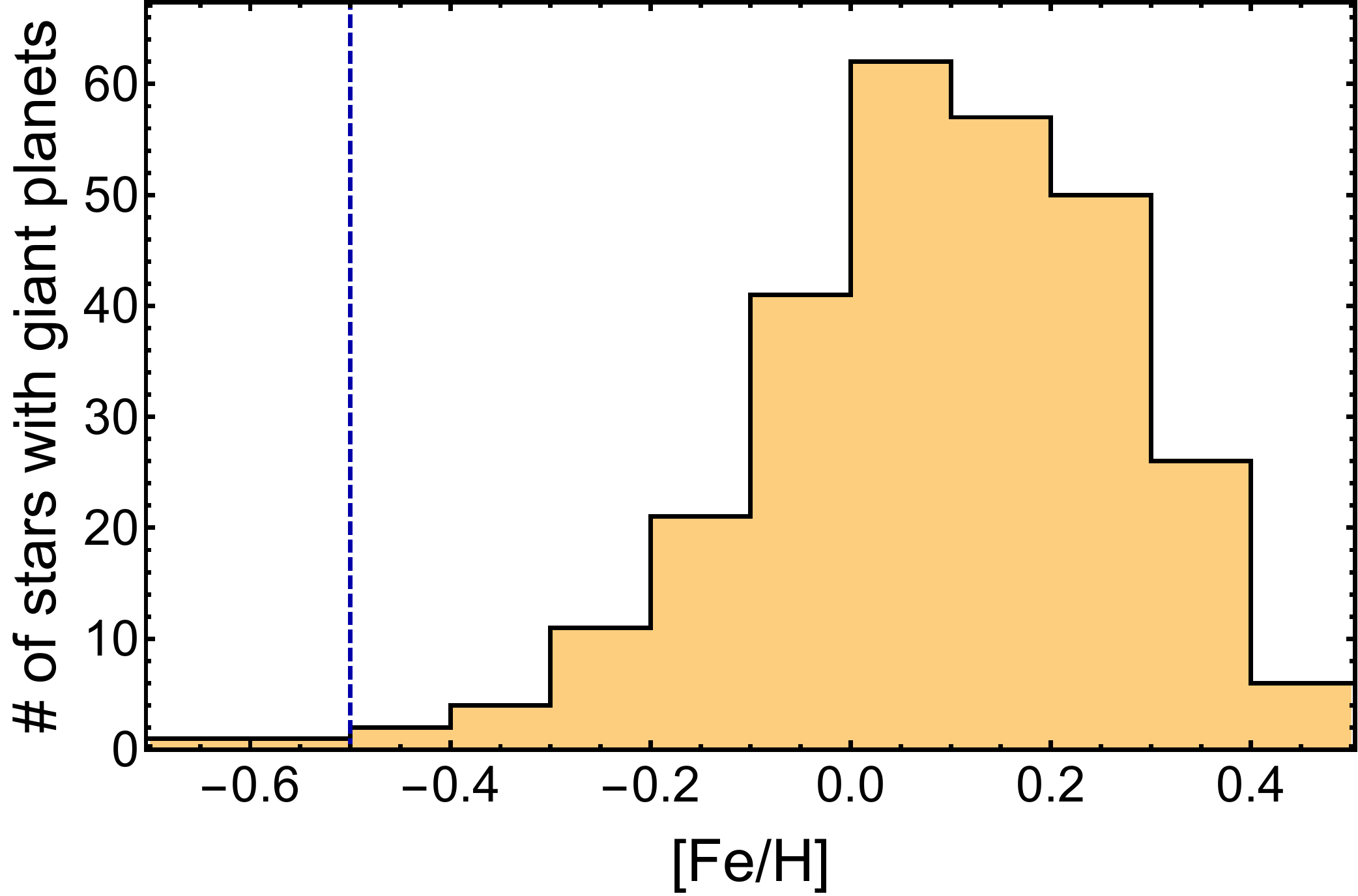}
  \caption{Frequency distribution of parent-star metallicity for transiting exoplanets with $M_{\rm p}>0.3\,M_{\rm Jup}$ \citep{hatzes:2015}. Data taken on 2016/05/09 from the Transiting Extrasolar Planet Catalogue (TEPCat; \citealt{southworth:2011}), which is available at http://www.astro.keele.ac.uk/jkt/tepcat/. The dashed vertical line refers to the threshold expected by the core-accretion theory \citep{mordasini:2012}.}
  \label{fig:metallicity}
\end{figure}

%%%%%%%%%%%%%%%%%%%%%%%%%%%
More than twenty years since their discovery, hot-Jupiter planets are still peculiar astrophysical objects, whose diversity, formation process(es) and migration mechanism(s) are not well understood. An accurate knowledge of their characteristics is mandatory to put precise constraints on theoretical models, many of which were proposed during the last three decades to explain them, and discriminate which best reproduce observations. In this context, transiting hot Jupiters are the most interesting to consider for detailed follow-up observations since their main physical parameters (mass, radius, equilibrium temperature, orbital eccentricity and obliquity, etc.) are fully measurable with various `standard' observational techniques.
Finding evidence of possible correlations between their properties and initial conditions can be useful for constructing a reasonable population of synthetic planets able to explain their formation and evolution. So far, the most convincing correlation is that between the stellar metallicity and the frequency of giant planets, which were mostly detected around metal-rich host stars \citep{gonzalez:1997,santos:2004,valenti:2005}, see Fig.\,\ref{fig:metallicity}. Besides having observational support, the idea that Jupiter planets are much more common around metal-rich stars also follows naturally from the core-accretion theory of planet formation, because in such environments a massive core can form more quickly, before the dispersal of the gas disc, and then accrete gas from the disc \citep{ida:2004,mordasini:2009}.

Moreover, transiting hot Jupiters are particularly well-suited for studying the atmospheres of this class of exotic giant planets. Indeed, since most of them are exceptionally bloated (probably caused by the intense irradiation received from their parent stars, e.g.\ \citealp{weiss:2013}), their relatively large sizes can give deep transits (typically $0.5$ to $3.5\%$). The transmission-spectroscopy technique has been widely used in recent years for probing the chemical composition at the day-night terminator of the atmosphere of these planets. In particular, the most convincing results have been obtained from the space with the Hubble and Spitzer space telescopes (e.g.\ \citealp{sing:2016}).

In this framework, we have for the last eight years run a large observational program with an array of medium-class telescopes, located in both the northern and southern hemispheres, with the aim of characterising a large sample of the planetary systems known to host transiting hot Jupiters \citep{mancinisouth:2016}. Our program is based on high-quality photometric follow-up observations of complete transit events by means of the telescope-defocussing technique. Importantly, all the datasets are reduced and analysed in a homogeneous way; different approaches in data analysis can indeed produce inhomogeneous results, which are then not straightforward to compare. Some of the telescopes that we are using are equipped with multi-band imaging cameras, which allows us to monitor transits simultaneously in different filters, enabling the possibility of probing the atmosphere of a transiting planet by measuring its transmission spectrum, similarly to transmission spectroscopy but with a lower spectral resolution.

In this work we present our results concerning the planetary system WASP-98, discovered by \citet{hellier:2014} within the SuperWASP survey \citep{pollacco:2006}. It hosts a low-density hot Jupiter (WASP-98\,b; $M_{\rm b} \approx 0.8 \, M_{\rm Jup}$; $R_{\rm b} \approx 1.1 \, R_{\rm Jup}$; $T_{\rm eq} \approx 1180$\,K), orbiting its parent star with a period of $\approx 2.96$\,days. The star is of spectral type G7\,V, brightness $V=13.0$\,mag, mass $M_{\rm A} \approx 0.7 \, M_{\sun}$ and radius $R_{\rm A} \approx 0.7 \, R_{\sun}$.

Interestingly, \citet{hellier:2014} found that WASP-98\,A is a very metal poor star, ${\rm [Fe/H]} =-0.60 \pm 0.19$, which is unusual for stars hosting hot Jupiters; another emblematic case is WASP-112\,b (${\rm [Fe/H]} =-0.64 \pm 0.15$, \citealt{anderson:2014}). Only very few planets in the core-accretion paradigm are expected to form at such low metallicities \citep{matsuo:2007,johnson:2012,mordasini:2012}. These would have required a high surface density of solid material and, therefore, a high disk mass to compensate a low dust to gas ratio. Alternatively, other planet formation paradigms, such as formation by gravitational instability in which planet frequency is independent of metallicity \citep{boss:2002}, may be able to explain the formation of these planets.
%We do not expect to find gas giants below the ${\rm [Fe/H]} \approx-0.5$ threshold \citep{johnson:2012,mordasini:2012}. The existence of WASP-98\,b, as well as that of %WASP-112\,b (${\rm [Fe/H]} =-0.64 \pm 0.15$, \citealt{anderson:2014}), therefore does not support the core-accretion theory for the formation of giant planets, instead %favouring the gravitational-instability scenario in which planet frequency is independent of metallicity \citep{boss:2002}. 
However, \citet{hellier:2014} stressed that their spectral analysis depended on spectra of a low signal-to-noise-ratio (S/N) and suggested to take higher S/N spectra for confirmation of such a low metallicity.

The low metallicity of WASP-98\,A encouraged us to make this planetary system a high-priority target for our photometric follow-up programme. Indeed, this paper concerns observations of new transit events of the planet and a more accurate spectroscopic study of the parent star. Thanks to new photometric and spectroscopic data collected at the MPG 2.2\,m telescope, we revised the main physical properties of the system and probed the terminator of the atmosphere of WASP-98\,b.
% (a similar case is WASP-112).

The paper is structured as follows. In Section~\ref{sec:observ-data-reduct} we describe the observations carried out and the data reduction process. Section~\ref{sec:light-curve-analysis} is devoted to the analysis of the photometric data and the refinement of the orbital ephemeris.  The revision of the physical parameters of the planetary system is discussed in Section~\ref{sec:physical-properties}, while in Section \ref{sec:radius-variation} we present a novel study of the variation of the radius of WASP-98\,b with wavelength. The results of this work are summarised in Section~\ref{sec:summary-conclusions}.

%%%%%%%%%%%%%%%%%%%%%%%%%%%
\section{Observations and data reduction}
\label{sec:observ-data-reduct}
%%%%%%%%%%%%%%%%%%%%%%%%%%%

\subsection{Photometry}

\begin{table*}
  \centering
  \caption{Details of the transit observations presented in this work, all
    carried out at the MPG~\(2.2~\mathrm{m}\) telescope located at ESO
    Observatory in La Silla, Chile.  `Date of first obs.' is the date when the
    first observation started, \(N_{\textup{obs}}\) is the number of
    observations, \(T_{\textup{exp}}\) is the exposure time,
    \(T_{\textup{obs}}\) is the observational cadence, and `Moon illum.' is the
    fractional illumination of the Moon at the midpoint of the transit.  The
    aperture sizes define the radii of the software apertures for the star, inner
    sky and outer sky, respectively.  Scatter is the r.m.s.\ scatter of the data
    versus a fitted model.}
  \label{tab:observations}
  \begin{tabular}{ccccccccccc}
    \hline
    Date of     & Start time & End time & $N_{\rm obs}$ & $T_{\rm exp}$ & $T_{\rm obs}$ & Filter       & Airmass           & Moon   & Aperture               & Scatter \\
    first obs.  & (UT)       & (UT)     &               & (s)           & (s)           &              &                   & illum. & radii (px)             & (mmag)  \\
    \hline
    2014 Oct 13 & 02:35      & 07:38    & 118           & \(110\)       & 154           & Sloan \(g'\) & \(1.85 \to 1.01\) & 74\%   & \(23\), \(50\), \(70\) & 0.77    \\
    2014 Oct 13 & 02:35      & 07:38    & 114           & \(110\)       & 154           & Sloan \(r'\) & \(1.85 \to 1.01\) & 74\%   & \(20\), \(50\), \(70\) & 0.81    \\
    2014 Oct 13 & 02:35      & 07:38    & 114           & \(110\)       & 154           & Sloan \(i'\) & \(1.85 \to 1.01\) & 74\%   & \(25\), \(50\), \(70\) & 0.68    \\
    2014 Oct 13 & 02:35      & 07:38    & 114           & \(110\)       & 154           & Sloan \(z'\) & \(1.85 \to 1.01\) & 74\%   & \(25\), \(50\), \(70\) & 0.79    \\[2pt]
    2015 Jan 07 & 01:09      & 06:02    & 115           & \(110\)       & 154           & Sloan \(g'\) & \(1.00 \to 1.85\) & 96\%   & \(23\), \(50\), \(70\) & 0.83    \\
    2015 Jan 07 & 01:09      & 06:02    & 115           & \(110\)       & 154           & Sloan \(r'\) & \(1.00 \to 1.85\) & 96\%   & \(25\), \(50\), \(70\) & 1.00    \\
    2015 Jan 07 & 01:09      & 06:02    & 113           & \(110\)       & 154           & Sloan \(i'\) & \(1.00 \to 1.85\) & 96\%   & \(25\), \(50\), \(70\) & 0.75    \\
    2015 Jan 07 & 01:09      & 06:02    & 114           & \(110\)       & 154           & Sloan \(z'\) & \(1.00 \to 1.85\) & 96\%   & \(25\), \(50\), \(70\) & 0.98    \\
    \hline
  \end{tabular}
\end{table*}

In October 2014 and January 2015, we observed two transit events of WASP-98\,b using the ``Gamma Ray Burst Optical and Near-Infrared Detector'' instrument (GROND; \citealt{greiner:2008}), which is hosted in the Cound\'{e}-like focus of the MPG 2.2\,m telescope, located at ESO La Silla, Chile. GROND is a seven-channel imager capable of performing simultaneous observations in four optical bands (\(g'\), \(r'\), \(i'\), \(z'\), similar to the Sloan filters) and three near-infrared bands (\(J\), \(H\), \(K\)). This instrument was primarily designed for rapid observations of the optical and near-infrared counterparts of gamma-ray bursts, but in recent years it has provided a very good performance in obtaining simultaneous, multi-colour light curves of planetary-transit events (e.g.\ \citealt{mancini:2014a,southworth:2015,ciceri:2016}). Dichroic elements are used to split the incoming stellar light into different paths leading to two sets of cameras.  For the two transits of WASP-98\,b we only used the optical channels, whose cameras are four back-illuminated $2048\times2048$ pixel E2V CCDs. The pixel size of each CCD is $13.5\,\mu$m and the focal length of the MPG 2.2m telescope is 17.6\,m, giving a plate scale of 0.158\,arcsec per pixel and field-of-view of \(5.4' \times 5.4'\).

During the two transits observations of WASP-98\,b, the telescope was autoguided and defocussed in order to reduce the effects of systematic-error sources and decrease the amount of time lost to CCD readout. This yielded high-quality photometric data, with a scatter $\leq 1$\,mmag in all the eight light curves (two transits each in four filters). The details of the photometric observations are reported in Table~\ref{tab:observations}.

A suitable number of calibration frames, bias and (sky) flat-field images, were taken on the same day as the observations. Master bias and flat-field images were created by median-combining all the individual bias and flat-field images, and used to calibrate the scientific images. Image motion was tracked by cross-correlating individual images with a reference one.  We then used the {\large {\sc daophot}} aperture-photometry algorithm \citep{stetson:1987} and the {\large {\sc defot}} package \citep{southworth:2009,southworth:2014} to extract the photometry of the target star from the calibrated images. Using the {\large {\sc aper}} routine\footnote{{\sc aper} is part of the {\large {\sc astrolib}} subroutine library distributed by NASA.}, we measured the flux and the error of the target and of four comparison stars in the field-of-view, selecting those of similar brightness to the target and not showing any significant brightness variation due to intrinsic variability or instrumental effects.  For each dataset, we tried different aperture size for both target and sky, and the final ones that we selected, reported in Table~\ref{tab:observations}, were those that gave the lowest scatter in the out-of-transit region. Light curves were then obtained by performing differential photometry using the reference stars in order to correct for non-intrinsic variations of the flux of the target, which are caused by atmospheric or airmass changes.

\begin{figure*}
  \centering
  \includegraphics[width=\textwidth]{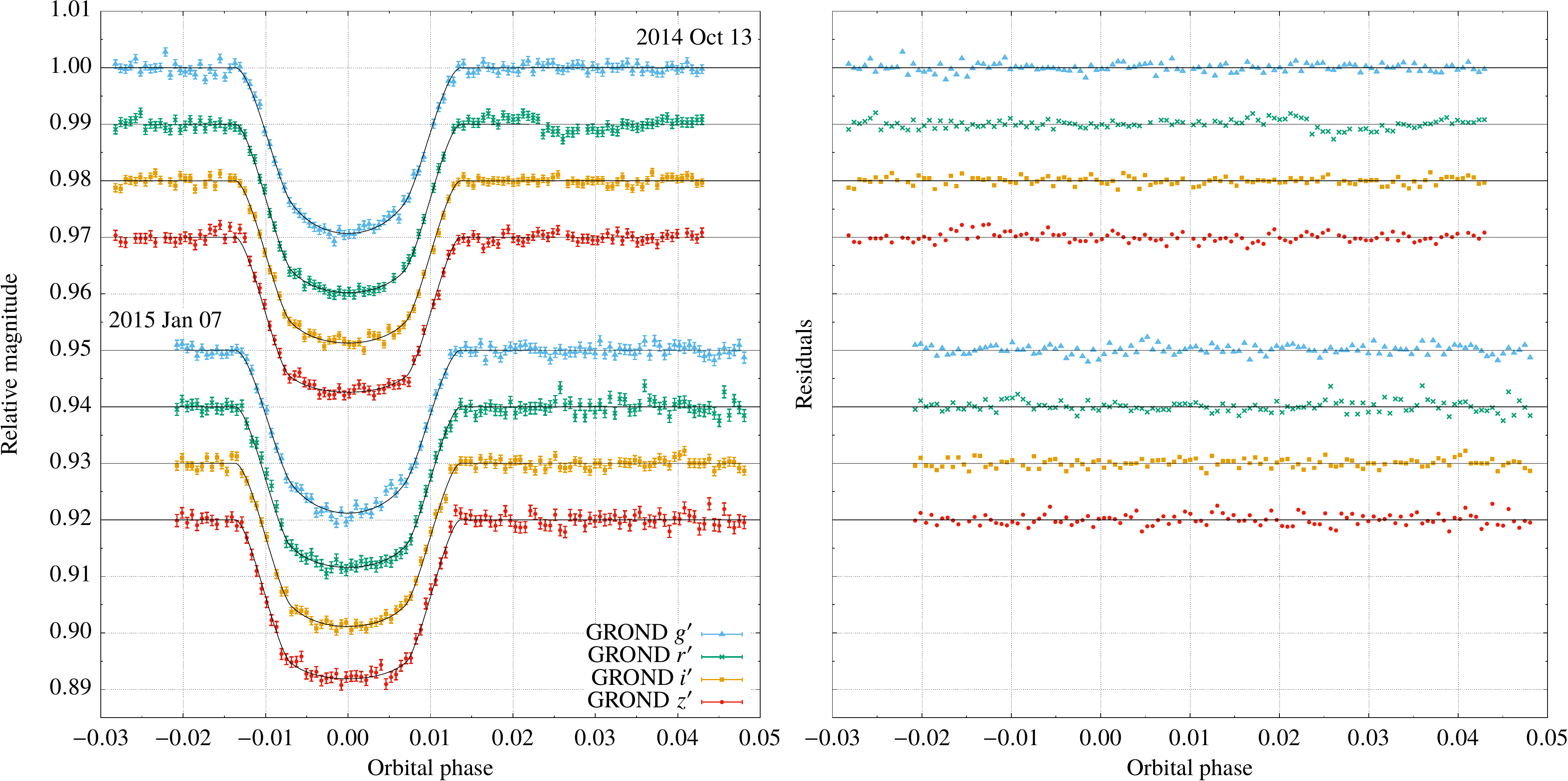}
  \caption{\emph{Left}: phased light curves of WASP-98 for the two observed
    transits in the four optical bands of GROND.  Each light curve is shown with its
    best fit obtained with \textsc{jktebop}.  The polynomial baseline functions
    have been removed from the data before plotting.  Labels indicate the date
    of the two observing runs.  The curves are shifted along the abscissae for
    clarity. \emph{Right}: the residuals relative to each fit.}
  \label{fig:light-curves}
\end{figure*}

The instrumental magnitudes were then transformed to differential-magnitude light curves, normalised to zero magnitude outside transit using second-order polynomials fitted to the out-of-transit data. The differential magnitudes are relative to a weighted combination of the four comparison stars.  The reference star weights and polynomial coefficients were simultaneously optimised to minimise the scatter in the out-of-transit data. The reader can inspect the final light curves in Fig.\,\ref{fig:light-curves}. For accurate timing of the event, the timestamps for the datapoints were converted to the BJD\,(TDB) timescale (Barycentric Julian Date in the Barycentric Dynamical Time, see~\citealt{eastman:2010}).

%%%%%%%%%%%%%%%%%%%%%%%%%%%
\subsection{Spectroscopy}
\label{sec:spectroscopy}
%%%%%%%%%%%%%%%%%%%%%%%%%%%
The spectral characteristics of the host star WASP-98\,A in the discovery paper \citep{hellier:2014} were based on spectra of relatively low S/N. We therefore acquired a sequence of spectra of the system on November 2, 2015, using the fibre-fed optical \'echelle spectrograph FEROS \citep{kaufer:1998}, also mounted on the MPG 2.2\,m telescope. FEROS covers a wide wavelength range ($370-860$\,nm) and has a mean resolving power of \(R = 48\,000 \pm 4000\).  The sequence consisted of three observations, each with an exposure time of 30\,min, taken without changing telescope's pointing.

The FEROS data were reduced using an automated pipeline that was built from a modular code that is able to generate pipelines for the processing and analysis of \'echelle spectra originated from different instruments (see \citealt{jordan:2014}). This pipeline ($i$) performs an optical extraction, following \citet{marsh:1989}, ($ii$) estimates the global wavelength solution of the \'echelle by fitting a two-dimensional Chebyshev polynomial to the position of the ThAr emission lines, ($iii$) computes the instrumental velocity shift during the night, by using the global wavelength solution and the comparison fibre, and ($iv$) computes the radial velocity (RV) with the cross-correlation function technique using a G2 mask that resembles the spectrum of a G2-type star. The individual FEROS spectra have a S/N of $\sim$45 per resolution element at 5150\,\AA.

The three spectra were median-combined, obtaining a spectrum with S/N$\sim$70 that was analysed using the Zonal Atmospherical Stellar Parameter Estimator (ZASPE) code \citep{brahm:2015}. ZASPE estimates the atmospheric stellar parameters of a star by using a grid of synthetic spectra and searching for the one that produces the lowest $\chi^2$ value. The comparison with the models is performed in regions of the spectra which are most sensitive regions to changes in the atmospheric parameters. These particular spectral zones are computed on-the-fly by the code by determining the mean gradient of the synthetic grid as function of the three atmospheric parameters. ZASPE uses an iterative algorithm in which, after each iteration, the estimated $v\sin{i}$ value of the star is updated by using the width of the cross-correlation function computed between the observed spectrum and the synthetic one with the atmospheric parameters obtained in the previous iteration.

The errors in the atmospheric parameters include the systematic effects produced by the imperfect modelling of the synthetic grid, and are obtained using a Monte Carlo procedure in which the optimal set of atmospheric parameters is determined in several ($\sim$100) realizations, where the depth of the absorption lines of the synthetic spectra are randomly modified according to the probability distribution function of mismatch factors. The probability distribution function is obtained by computing the mismatch factors between the data and the optimal synthetic spectrum in each of the sensitive spectral regions. The synthetic grid used in this case was generated with the {\sc spectrum} code \citep{gray:1999} with the line list provided by the code. However, the oscillator strengths of several ($\sim$300) prominent lines were empirically calibrated using a set of stars with interferometrically and asteroseismologically derived parameters.
The results coming from the ZASPE analysis are discussed in Sect.\,\ref{sec:physical-properties}.

%%%%%%%%%%%%%%%%%%%%%%%%%%%
\section{Light curve analysis}
\label{sec:light-curve-analysis}
%%%%%%%%%%%%%%%%%%%%%%%%%%%

%%%%%%%%%%%%%%%%%%%%%%%%%%%%%%%%%%%%%%%%%%%%%%%%%%%%%%%%
\subsection{Photometric parameters}
\label{sec:photom-parameters}
%%%%%%%%%%%%%%%%%%%%%%%%%%%%%%%%%%%%%%%%%%%%%%%%%%%%%%%%

We separately fitted each of the eight GROND light curves with the {\large {\sc jktebop}}\footnote{{\large {\sc jktebop}} is written in FORTRAN 77 and its source code is available at \url{http://www.astro.keele.ac.uk/jkt/codes/jktebop.html}.} program \citep{southworth:2013}. The parameters that we fitted were: the orbital period and inclination ($P$ and $i$), the time of transit midpoint ($T_{0}$), and the sum and ratio of the fractional radii\footnote{The fractional radii are defined as $r_{\mathrm{A}} = R_{\mathrm{A}}/a$ and $r_{\mathrm{b}} = R_{\mathrm{b}}/a$, where $R_{\mathrm{A}}$ and $R_{\mathrm{b}}$ are the true radii of the star and planet, and $a$ is the semi-major axis.} ($r_{\mathrm{A}}+r_{\mathrm{b}}$ and $k=r_{\mathrm{b}}/r_{\mathrm{A}}$).  We also modelled the limb darkening (LD) of the star to properly constrain the planetary system's quantities; we applied a quadratic law for describing this effect and used the LD coefficients provided by the stellar models of \citet{claret:2004}. Since the two LD coefficients are very strongly correlated, we chose to fit for only one of them. Thus, we included the linear coefficient as a free parameter and we keep fixed the non-linear one, which was only perturbed by \(\pm 0.1\) on a flat distribution during the error-analysis process. The coefficients of a second-order polynomial were also fitted in order to fully account for the uncertainty in the detrending of the light curves. We fixed the orbital eccentricity to zero, based on the results of \citet{hellier:2014}.  Finally, we also considered that the uncertainties assigned by {\large {\sc aper}} to each photometric point tend to be underestimated, as this routine does not take into account correlated (red) noise, which generally afflicts time-series photometry \citep{carter:2009}. We therefore inflated the error bars to give a reduced $\chi^{2}$ of $\chi^{2}_{\nu}=1$ in the {\large {\sc jktebop}} best-fit process for each light curve.

Thanks to {\large {\sc jktebop}}, the best-fitting values for each parameter were obtained through a Levenberg--Marquardt minimisation, while uncertainties were estimated by running both a Monte Carlo and a residual-permutation algorithm \citep{southworth:2008}.  In particular, \(10\,000\) simulations for both the algorithms were run, and the largest of the two \(1\sigma\) values were adopted as the final uncertainty for each parameter.

The results of the fits of each light curve are summarised in Table~\ref{tab:parameters}, which also shows the final values of each parameter, obtained by means of a weighted average of the values extracted from the fit of all the individual light curves, using the relative uncertainties as a weight.  The photometric parameters found by \citet{hellier:2014} are reported in the same table for comparison; they are in good agreement with our results, but are much less precise.

\begin{table*}
  \centering
  \caption{Parameters of the fit to the light curves of WASP-98 from the
    \textsc{jktebop} analysis. The final parameters are given in bold and
    the parameters found by \citet{hellier:2014} are reported below for comparison.}
  \label{tab:parameters}
  \begin{tabular}{lccccc}
    \hline
    Source                     & \(r_{\textup{A}} + r_{\textup{b}}\)  & \(k\)                                & \(i(\degr)\)                  & \(r_{\textup{A}}\)                   & \(r_{\textup{b}}\)                   \\
    \hline
    GROND \(g'\) (2014 Oct 13) & \(0.1064 \pm 0.0028\)                & \(0.1615 \pm 0.0047\)                & \(86.35 \pm 0.25\)            & \(0.0916 \pm 0.0022\)                & \(0.014\,79 \pm 0.000\,72\)          \\
    GROND \(r'\) (2014 Oct 13) & \(0.1047 \pm 0.0027\)                & \(0.1590 \pm 0.0037\)                & \(86.47 \pm 0.22\)            & \(0.0904 \pm 0.0021\)                & \(0.014\,37 \pm 0.000\,60\)          \\
    GROND \(i'\) (2014 Oct 13) & \(0.1078 \pm 0.0021\)                & \(0.1599 \pm 0.0033\)                & \(86.28 \pm 0.16\)            & \(0.0929 \pm 0.0017\)                & \(0.014\,86 \pm 0.000\,49\)          \\
    GROND \(z'\) (2014 Oct 13) & \(0.1055 \pm 0.0021\)                & \(0.1583 \pm 0.0033\)                & \(86.44 \pm 0.17\)            & \(0.0911 \pm 0.0017\)                & \(0.014\,42 \pm 0.000\,47\)          \\[2pt]
    GROND \(g'\) (2015 Jan 07) & \(0.1035 \pm 0.0025\)                & \(0.1563 \pm 0.0025\)                & \(86.54 \pm 0.21\)            & \(0.0895 \pm 0.0020\)                & \(0.013\,99 \pm 0.000\,51\)          \\
    GROND \(r'\) (2014 Jan 07) & \(0.1070 \pm 0.0027\)                & \(0.1570 \pm 0.0017\)                & \(86.19 \pm 0.23\)            & \(0.0924 \pm 0.0023\)                & \(0.014\,51 \pm 0.000\,48\)          \\
    GROND \(i'\) (2014 Jan 07) & \(0.1082 \pm 0.0025\)                & \(0.1602 \pm 0.0018\)                & \(86.27 \pm 0.20\)            & \(0.0932 \pm 0.0021\)                & \(0.014\,94 \pm 0.000\,47\)          \\
    GROND \(z'\) (2014 Jan 07) & \(0.1049 \pm 0.0030\)                & \(0.1571 \pm 0.0015\)                & \(86.49 \pm 0.25\)            & \(0.0906 \pm 0.0025\)                & \(0.014\,24 \pm 0.000\,49\)          \\
    \hline
    \textbf{Final results}     & \(\mathbf{0.106\,13 \pm 0.000\,87}\) & \(\mathbf{0.158\,24 \pm 0.000\,81}\) & \(\mathbf{86.38 \pm 0.07}\) & \(\mathbf{0.091\,58 \pm 0.000\,70}\) & \(\mathbf{0.014\,53 \pm 0.000\,18}\) \\
    \hline
    \cite{hellier:2014} & \(0.1051 \pm 0.0060\)                & \(0.161 \pm 0.010\)                  & \(86.3 \pm 0.1\)              & \(0.0905 \pm 0.0051\)                & \(0.014\,60 \pm 0.000\,94\)          \\
    \hline
  \end{tabular}
\end{table*}

%%%%%%%%%%%%%%%%%%%%%%%%%%%
\subsection{New orbital ephemeris}
\label{sec:new-ephemeris}
%%%%%%%%%%%%%%%%%%%%%%%%%%%

\begin{table}
  \centering
  \caption{Mid-transit times of WASP-98\,b and their residuals to the fitted model. The numbers in brackets indicate the uncertainties on the last digit of the preceding number.} %
  \label{tab:mid-transit}
  \resizebox{\columnwidth}{!}{%
    \begin{tabular}{lrrl}
      \hline
      Time of minimum                               & Epoch & Residual\,~~~      & ~~~~~~~~~ Source                     \\
      BJD\,(TDB)\,$-2\,400\,000$ &       & (JD)\,~~~~~~         &                            \\
      \hline
      \(56\,333.39207(10)\)                       &     0  & \(-0.000005\) & ~~~~\citet{hellier:2014}  \\[2pt]
      \(56\,943.69613(23)\)                       & 206   & \( 0.000141\) & This work (GROND \(g'\))   \\
      \(56\,943.69608(19)\)                       & 206   & \( 0.000091\) & This work (GROND \(r'\))   \\
      \(56\,943.69600(18)\)                       & 206   & \( 0.000011\) & This work (GROND \(i'\))   \\
      \(56\,943.69594(22)\)                       & 206   & \(-0.000049\) & This work (GROND \(z'\))   \\[2pt]
      \(57\,029.61255(14)\)                       & 235   & \(-0.000010\) & This work (GROND \(g'\))   \\
      \(57\,029.61277(16)\)                       & 235   & \( 0.000210\) & This work (GROND \(r'\))   \\
      \(57\,029.61250(15)\)                       & 235   & \(-0.000060\) & This work (GROND \(i'\))   \\
      \(57\,029.61232(16)\)                       & 235   & \(-0.000240\) & This work (GROND \(z'\))   \\
%      \(56\,333.39207(10)\)                       & 0  & \(-0.0000057\) & ~~~~\citet{hellier:2014}  \\[2pt]
%      \(56\,943.69613(23)\)                       & 206   & \( 0.0001420\) & This work (GROND \(g'\))   \\
%      \(56\,943.69608(19)\)                       & 206   & \( 0.0000920\) & This work (GROND \(r'\))   \\
%      \(56\,943.69600(18)\)                       & 206   & \( 0.0000120\) & This work (GROND \(i'\))   \\
%      \(56\,943.69594(22)\)                       & 206   & \(-0.0000480\) & This work (GROND \(z'\))   \\[2pt]
%      \(57\,029.61255(14)\)                       & 235   & \(-0.0000082\) & This work (GROND \(g'\))   \\
%      \(57\,029.61277(16)\)                       & 235   & \( 0.0002118\) & This work (GROND \(r'\))   \\
%      \(57\,029.61250(15)\)                       & 235   & \(-0.0000582\) & This work (GROND \(i'\))   \\
%      \(57\,029.61232(16)\)                       & 235   & \(-0.0002382\) & This work (GROND \(z'\))   \\
      \hline
    \end{tabular}}
\end{table}
%
%\begin{figure*}
%  \centering
%  \includegraphics[width=\textwidth]{period}
%  \caption{Plot of the residuals of the timing of mid-transit of WASP-98 versus
%    a linear ephemeris.  See also Table~\ref{tab:mid-transit}.}
%  \label{fig:period}
%\end{figure*}

One of the parameters obtained by fitting planetary-transit light curves with {\large {\sc jktebop}} is the time of mid transit. We can therefore use the values of $T_{0}$ for each of the new eight light curves to refine the ephemeris of the WASP-98\,b transits. In doing this, we also included the value of the epoch of mid-transit time from the discovery paper, after having converted it to BJD\,(TDB). We performed a weighted linear least-squares fit to all the mid-transit times versus their cycle number (see Table~\ref{tab:mid-transit}), using the Levenberg--Marquardt algorithm, obtaining
\begin{equation}
  T_{0} = \textup{BJD}\,({\rm TDB})~2\,456\,333.392075(82) + 2.96264036(43)\,E,
\end{equation}
where the numbers in brackets indicate the uncertainties on the last digits of the preceding number, and \(E\) is the number of orbits that the planet has completed since the reference time. The \(\chi_{\nu}^{2}\) of the fit is \(0.68\), indicating an unusually good agreement between the linear ephemeris and the observations. However, given the small number of the observed transit events of WASP-98\,b, it is not possible to draw any useful conclusions about the possible presence of transit timing variations.

%%%%%%%%%%%%%%%%%%%%%%%%%%%
\section{Physical properties}
\label{sec:physical-properties}
%%%%%%%%%%%%%%%%%%%%%%%%%%%
As in previous papers of our series, we redetermined the physical properties of the WASP-98 system following the \emph{Homogeneous Studies} approach (see \citealt{southworth:2012} and references therein), which consists of combining the measured parameters from the light curves, spectroscopic observations, and constraints on the properties of the host star from theoretical stellar evolutionary models. The spectroscopic properties of the host star were obtained from the analysis of the FEROS spectra performed with ZASPE, Sect.\,\ref{sec:spectroscopy}, which returned values for the following quantities: effective temperature $T_{\rm eff}$, logarithmic surface gravity $\log{g_{\star}}$, metallicity, and projected rotational velocity $v \sin{i_{\star}}$, where $i_{\star}$ is the inclination of the stellar rotation axis with respect to the line of sight. We also utilised the velocity amplitude of the RV variation, $K_{\mathrm{A}}=150 \pm 10$\,m\,s$^{-1}$, which was taken from \citet{hellier:2014}.

The input parameters were used to determine the physical properties of the system, using the {\large {\sc jktabsdim}} code \citep{southworth:2009a}. This was done by iteratively modifying the velocity amplitude of the planet, $K_{\rm b}$, to maximise the agreement between the measured $R_{\mathrm{A}}/a$ and $T_{\mathrm{eff}}$ and those predicted by a set of theoretical models. A wide range of possible ages for the host star was considered, as was each of five sets of theoretical models, from which we obtained five different estimates of the stellar logarithmic surface gravity. We took the unweighted mean of these and used this new value of $\log{g_{\star}}$ fixed for a second iteration with ZASPE \citep{brahm:2015,mancini:2015}. We thus determined new spectral parameters: $T_{\rm eff}=5473 \pm 121$\,K, [Fe/H] $=-0.49 \pm 0.10$\,dex, $v \sin{i_{\star}}=0.99 \pm 1.08$\,km\,s$^{-1}$, which were then used to run {\large {\sc jktabsdim}} for a second time to obtain final values of the output parameters. Each output parameter had five different estimates, obtained using the five theoretical stellar models, and the unweighted mean was taken as the final value of the parameter. Its systematic error was obtained as the maximum deviation between the final parameter value and the five individual values from the different theoretical models. Our final values are reported in Table\,\ref{tab:derived-params} and are more precise compared to those from the discovery paper \citep{hellier:2014}. In particular, we found that the masses of both the star and the planet are significantly larger than previously measured, by $17\%$ and $11\%$ respectively.
\begin{table*}
\centering
\caption{Derived physical properties of the WASP-98 planetary system.  The values found by \citet{hellier:2014} are given for comparison. Those parameters, which have a dependence on stellar theory, have two uncertainties: the first error bar is the statistical error, which depends on measured spectroscopic and photometric parameters; the second is the systematic error, coming from the use of theoretical stellar models. \newline
\(^{(*)}\) This is the stellar age obtained from lithium abundance measurements. \(^{(\dagger)}\) This is the stellar age obtained from gyrochronology.}
  \label{tab:derived-params}
  \begin{tabular}{lcccc}
    \hline
    Parameter                     & Symbol                     & Unit                            & This work                   & \citet{hellier:2014}   \\ %
    \hline
    \emph{Stellar parameters}     &                            &                                 &                                           &                                       \\
    Mass                          & \(M_{\textup{A}}^{}\)      & \(M_{\sun}^{}\)                 & \(0.809 \pm 0.053 \pm 0.036\)             & \(0.69 \pm 0.06\)                     \\
    Radius                        & \(R_{\textup{A}}^{}\)      & \(R_{\sun}^{}\)                 & \(0.741 \pm 0.018 \pm 0.011\)             & \(0.70 \pm 0.02\)                     \\
    Logarithmic surface gravity   & \(\log g_{\textup{A}}^{}\) & cgs                             & \(4.606 \pm 0.012 \pm 0.006\)             & \(4.583 \pm 0.014\)                   \\
    Density                       & \(\rho_{\textup{A}}^{}\)   & \(\rho_{\sun}^{}\)              & \(1.987 \pm 0.046\)                       & \(1.99 \pm 0.07\)                     \\
    Effective temperature         & \(T_{\textup{eff}}\)       & K                               & \(5473.0 \pm 121.0\)                      & \(5550 \pm 140\)                      \\
    Metallicity                   & [Fe/H]                     & dex                             & \(-0.49 \pm 0.10\)                        & \(-0.60 \pm 0.19\)                    \\
    Projected rotational velocity & \(v \sin i_{\star}\)       & \(\mathrm{km}~\mathrm{s}^{-1}\) & \(0.99 \pm 1.08\)                         & \(<0.5\)                              \\
    Age                           &                            & Gyr                             & \(2.7\,^{+6.2}_{-2.5}\,^{+2.9}_{-1.0}\)   & \(>3~^{(*)}\) | \(>8~^{(\dagger)}\)                         \\
    \hline
    \emph{Planetary parameters}   &                            &                                 &                                           &                                       \\
    Mass                          & \(M_{\textup{b}}^{}\)      & \(M_{\textup{Jup}}^{}\)         & \(0.922 \pm 0.075 \pm 0.027\)             & \(0.83 \pm 0.07\)                     \\
    Radius                        & \(R_{\textup{b}}^{}\)      & \(R_{\textup{Jup}}^{}\)         & \(1.144 \pm 0.029 \pm 0.017\)             & \(1.10 \pm 0.04\)                     \\
    Surface gravity               & \(g_{\textup{b}}^{}\)      & \(\mathrm{m}~\mathrm{s}^{-2}\)  & \(17.5 \pm 1.2\)                          & \(15.8 \pm 1.1\)                      \\
    Density                       & \(\rho_{\textup{b}}^{}\)   & \(\rho_{\textup{Jup}}^{}\)      & \(0.576 \pm 0.046 \pm 0.008\)             & \(0.63 \pm 0.06\)                     \\
    Equilibrium temperature       & \(T_{\textup{eq}}\)       & \(\mathrm{K}\)                  & \(1171 \pm26\)                            & \(1180 \pm 30\)                       \\
    Safronov number               & \(\Theta\)                 &                                 & \(0.0750 \pm 0.0054 \pm 0.0011\)          & ---                                   \\
    \hline
    \emph{Orbital parameters}     &                            &                                 &                                           &                                       \\
    Period                        & \(P\)                      & d                               & \(2.96264036 \pm 0.00000043\)     & \(2.9626400 \pm 0.0000013\)       \\
    Epoch of mid-transit          & \(T_{0}\)                  & BJD\,(TDB) & \(2\,456\,333.392075 \pm 0.000082\)   & \(2\,456\,333.39207 \pm 0.00010\) \\
    Semimajor axis                & \(a\)                      & AU                              & \(0.03762 \pm 0.00083 \pm 0.00055\) & \(0.036 \pm 0.001\)                   \\
    \hline
  \end{tabular}
\end{table*}

%%%%%%%%%%%%%%%%%%%%%%%%%%%
\section{Variation of the planetary radius with wavelength}
\label{sec:radius-variation}
%%%%%%%%%%%%%%%%%%%%%%%%%%%
An important role in determining the atmospheric properties of hot Jupiters is played by the amount of radiation that comes from parent stars. Its influence on the planetary surfaces causes different atmospheric chemical mixing ratios and opacities in planetary atmospheres. A distinction of two class of planets was initially proposed by \citet{fortney:2008}, who suggested to distinguish hot Jupiters as pM and pL planets depending on their equilibrium temperature, which translates into the possible presence or absence of strong absorbers, such as gaseous titanium oxide (TiO), in their upper atmospheres \citep{burrows:2008}.

The opacities of TiO should be extremely strong for the pM class, causing temperature inversions with increasing altitude and an observable variation of planetary radius with wavelength in the optical bands (between 450 and 700\,nm) during transit events. However, this hypothesis might be problematic and macroscopic mixing processes are required to prevent Ti rain-out from the upper altitudes of planetary atmospheres, maintaining thus a quite high abundance of this species \citep{spiegel:2009}. The existence of other optical absorbers at high altitudes, like vanadium oxide (VO) or sulphanyl (HS), has been speculated (e.g.\ \citealt{burrows:2007,fortney:2008,zahnle:2009}), but also the inhibition of oxide formation due to a carbon-to-oxygen (C/O) ratio $\geq 1$ could intervene in the formation of temperature inversions \citep{madhusudhan:2012}. Possible detections of TiO absorbing features were observed in very few cases, i.e.\ HAT-P-8\,b \citep{mancini:2013} and WASP-121 \citep{evans:2016} via broad-band photometric observations of transit events, and WASP-33\, \citep{haynes:2015} by spectroscopic time series observations of occultation events.

For the colder pL-class planets a significant contribution to the opacity in the optical band is expected from Na\,{\large {\sc i}}  at $\approx 590$\,nm, and K\,{\large {\sc i}} at $\approx 770$\,nm. Besides atomic and molecular absorption, another strong variation of opacity with wavelength could be caused by Rayleigh scattering in the blue region of the optical spectrum. All of these were actually observed in the transmission spectra of several transiting exoplanets with very good confidence (see e.g.\ \citealt{sing:2016}).

%The observation of absorption features in the transmission spectra of hot Jupiters can therefore be extremely useful for probing the chemical composition and dynamical %processes in their atmospheres.

\begin{table}
  \centering
  \caption{Values of the radii ratio \(k\) for each passband.}
  \label{tab:k-bands}
  \begin{tabular}{lccc}
    \hline
    Passband     & Central         & FWHM      & \(k\) \\
                 & wavelength (nm) & (nm)      &       \\
    \hline
    GROND \(g'\) & \(459.0\)  & \(137.9\) & $0.15826 \pm 0.00036 $ \\
    GROND \(r'\) & \(622.0\)   & \(138.2\) & $0.15859 \pm 0.00038$  \\
    GROND \(i'\) & \(764.0\)   & \(153.5\) & $0.15828 \pm 0.00034$  \\
    GROND \(z'\) & \(898.9\)  & \(137.0\) & $0.15600 \pm 0.00037$ \\
    \hline
  \end{tabular}
\end{table}
\begin{figure*}
  \centering
  \includegraphics[width=18cm]{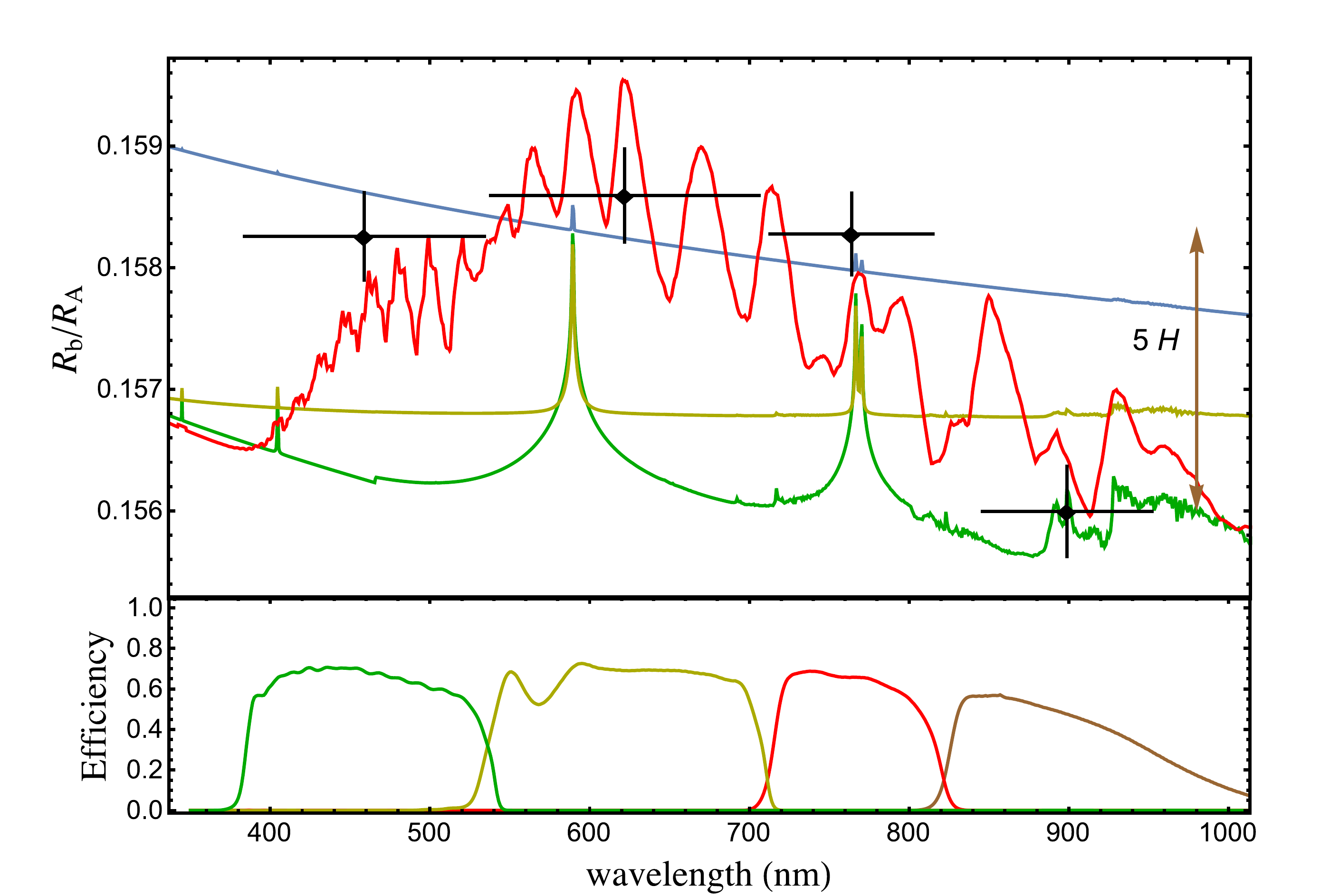}
  \caption{Variation of the ratio of the radii with wavelength. The black points are the weighted mean $k$ values
from the GROND observations. The vertical bars represent the relative uncertainties and the horizontal bars show
the FWHM transmission of the passbands. Synthetic spectra for WASP-98\,b, obtained with the \emph{petitCODE},
are shown as coloured lines. The green line refers to a clear atmosphere, the blue line is as the green one but with Rayleigh scattering increased by a factor of 1000 to mimic a very hazy atmosphere, the yellow line shows the case of a cloudy atmosphere, and the red line shows the case in which we consider the opacity contribution due to TiO and VO, but suppressing the condensation of Ti and V. Offsets are applied to the models to provide the best fit to our radius measurements. The atmospheres were computed for a planetary metallicity as that of the parent star. The size of five atmospheric pressure scale heights ($5\,H$) is shown on the right of the plot. Transmission curves of the GROND filters are shown in the bottom panel. }
  \label{fig:radiusvariation}
\end{figure*}
%

%%%%%%%%%%%%%%%%%%%%%%%%%%%
\subsection{Photometric transmission spectrum}
\label{sec:transmission-spectrum}
%%%%%%%%%%%%%%%%%%%%%%%%%%%
We took advantage of the ability of the GROND instrument, which provides simultaneous light curves at different colours, to investigate the possible variation of the planetary radius of WASP-98\,b, in terms of $k=r_{\rm b}/r_{\rm A}=R_{\rm b}/R_{\rm A}$, with wavelength. Following the approach of \citet{southworth:2012a}, we estimated the ratio of the radii in each of the four optical passband, fixing all the other photometric parameters to their best-fit values. In this way we obtained a set of $k$ values which are not affected by common sources of uncertainty and are directly comparable. They are reported in Table\,\ref{tab:k-bands}. Whilst we did not find any variation of the planetary radius among the $g^{\prime}$, $r^{\prime}$ and $i^{\prime}$ bands within the experimental uncertainties, we did detect a significant variation between these and the $z^{\prime}$ bands. The variation is roughly $5.5$ pressure scale heights\footnote{The pressure scale height is defined as $H = \frac{k_{\rm B}T_{\rm eq}}{\mu_{\mathrm{m}}\,g_{\mathrm{p}}}$, where $k_{\rm B}$ is Boltzmann's constant and $\mu_{\mathrm{m}}$ the mean molecular weight.} for $r^{\prime}$ versus $z^{\prime}$, with a confidence level of roughly 6$\sigma$ (see Fig.\,\ref{fig:radiusvariation}).

%%%%%%%%%%%%%%%%%%%%%%%%%%%
\subsection{The \emph{petitCODE}}
\label{sec:petitCODE}
%%%%%%%%%%%%%%%%%%%%%%%%%%%
We investigated the radius variation of WASP-98\,b by calculating reasonable synthetic transmission spectra and comparing them to the observed one. Self-consistent modelling of one-dimensional atmospheric structures and spectra were produced by making use of the \emph{petitCODE} \citep{molliere:2015}. As already mentioned in \citet{mancini:2016}, this code has recently been extended to also calculate transmission spectra, including Rayleigh scattering, TiO and VO opacities. Here, we briefly report on new capabilities of the code, which will be described in detail in a forthcoming paper (Molli{\`e}re et al., in prep.):

\begin{itemize}
\item {\it Clouds:} The code has been extended by a cloud module, for which the cloud model, as reported in \citet{ackerman:2001}, was used, including some minor changes partially to adapt it for the case of hot Jupiters. For the high altitude (low pressure) regions of the atmosphere, we introduced an eddy diffusion coefficient, which increases with decreasing pressure, as it has been found in general circulation model (GCM) simulations of hot Jupiters. The pressure dependence of this contribution is approximately proportional to $P^{-0.5}$ \citep{parmentier:2013,agundez:2014}. Further, above the convective region, we used a power-law decrease of the convective eddy diffusion coefficient, starting with the coefficient derived from mixing length theory in the last convective layer. This simulates convective overshoot in the deep, high pressure regions of the atmosphere \citep{ludwig:2002,helling:2008}.
We compared our results of the cloud mass density for self-luminous planets set up identically to those reported in \citet{ackerman:2001}, and found a very good agreement.
To calculate the cloud opacities for homogeneous spheres, we used cross-sections derived from Mie theory, calculated using a dust opacity code developed for distributions of hollow spheres (DHS) opacities \citep{min:2005}, which makes use of software reported in \citet{toon:1981}. At present, we include MgSiO$_3$, Mg$_2$SiO$_4$, Fe, KCl and Na$_2$S clouds with the real and complex parts of the refractive indices taken from \citet{jaeger:2003} for MgSiO$_3$ and Mg$_2$SiO$_4$, \citet{pollack:1994} for Fe, \citet{palik:2012} for KCl and \citet{morley:2002} for Na$_2$S.
\item {\it Scattering:} In the previous version of the code the effect of scattering was only included as an extinction source when including Rayleigh scattering during the computation of transmission spectra. To also include the effect of scattering of the clouds and molecules (via molecular Rayleigh scattering) on the temperature structure of the atmosphere, we implemented isotropic scattering for both the stellar insolation and the planetary flux. In order to speed up convergence, we made use of locally accelerated lambda iteration \citep[ALI, see][]{olson:1986} and Ng acceleration \citep{Ng:1974}.
\item {\it Line cutoff:} While our previous opacities were without any sub-Lorentzian line cutoff, we now include an exponential decrease for all molecular and atomic lines, because this is a more realistic scenario. We use the measurements by \citet{hartmann:2002} for all molecules but CO$_2$, and a fit to the CO$_2$ measurements of \citet{burch:1969} carried out by Bruno Bezard (priv.\ comm.) for CO$_2$. Note that the measurements by \citet{hartmann:2002} are made for CH$_4$ broadened by H$_2$. Due to the lack of measurements for other species we use the CH$_4$ measurements for other species as well, except for CO$_2$, for which dedicated measurements exist.
\end{itemize}

%%%%%%%%%%%%%%%%%%%%%%%%%%%
\subsection{Comparison between observations and models}
\label{sec:comparison}
%%%%%%%%%%%%%%%%%%%%%%%%%%%
We used our measured physical properties from Table\,\ref{tab:derived-params} for calculating self-consistent atmospheric structures of WASP-98\,b. The stellar irradiation was calculated assuming a global average, i.e.\ the flux received at the top of the atmosphere was a quarter of the flux at the substellar point. For the planet's enrichment we used the host star's metallicity ([Fe/H] $_{\rm b}=-0.49$). We considered the following five cases:
\begin{enumerate}
\item[$(i)$] {\it Cloud-free case}: Clear atmosphere
%at [Fe/H] $_{\rm b}=-0.49$
without TiO and VO opacities.
\item[$(ii)$] {\it Hazy case}: Clear atmosphere
%at [Fe/H] $_{\rm b}=-0.49$
with the H$_2$ Rayleigh scattering opacity multiplied by 1000 to mimic a strong haze-like absorber. No TiO and VO opacities. Here we did not calculate a dedicated atmospheric $P$-$T$ structure, but instead used the structure derived from the clear case.
\item[$(iii)$] {\it Cloudy case}: Cloudy atmosphere
%at [Fe/H] $_{\rm b}=-0.49$
without TiO and VO opacities. We considered MgSiO$_3$, Mg$_2$SiO$_4$, Fe, KCl and Na$_2$S clouds and used a settling factor $f_{\rm sed}=3$, where $f_{\rm sed}$ is the cloud particle mass averaged ratio between the particle settling and the atmospheric convective mixing velocity.
\item[$(iv)$] {\it Standard TiO/VO case}: Clear atmosphere,
%at [Fe/H] $_{\rm b}=-0.49$,
including TiO and VO opacities.
\item[$(v)$] {\it Suppressed Ti/V condensation case}: Clear atmosphere,
%at [Fe/H] $_{\rm b}=-0.49$,
including TiO and VO opacities and suppressing the condensation of Ti and V.
\end{enumerate}
A comparison of the GROND data and the synthetic spectra can be seen in Fig.\,\ref{fig:radiusvariation}.

It is evident that the data are not well fit by the cloud-free, hazy or cloudy cases when TiO and VO is neglected. By including TiO/VO, in the standard case, no differences to the cloud-free case can be seen, because Ti and V are condensed and therefore cannot form TiO and VO. Only in the case in which we suppress the condensation of Ti and V, in which gaseous TiO and VO can form, do we attain a reasonable fit between the observations and data. This is a curious result which should not be taken at face value, however. The nominal thermochemical model clearly forbids the existence of gaseous TiO and VO in an atmosphere such that of WASP-98\,b. An additional optical absorber seems to be required in this case. Potential candidates could be metal hydrides such as CaH, MgH, CrH, TiH and FeH which have lower condensation temperatures but for which WASP-98b is right on the brink, and potentially too cold, to allow for these molecules' existence in the gas phase \citep[see, e.g.][]{sharp:2007,visscher:2010}. Further, most of the metal hydrides listed above (except for perhaps MgH) have a spectral shape inconsistent with the GROND data \citep[see][]{sharp:2007}. Note that \emph{petitCODE} does currently not include metal hydride opacities and not all of the above mentioned metal hydrides are currently treated in its equilibrium chemistry network. Moreover, in the standard cloud-free case the atmosphere is too cold to have appreciable amounts of MgH present. New measurements with higher spectral resolution are clearly needed to shed light on the nature of the optical absorber in the atmosphere of WASP-98\,b.

%%%%%%%%%%%%%%%%%%%%%%%%%%%
\section{Summary and conclusions}
\label{sec:summary-conclusions}
%%%%%%%%%%%%%%%%%%%%%%%%%%%

\begin{figure}
\centering
 \includegraphics[width=\columnwidth]{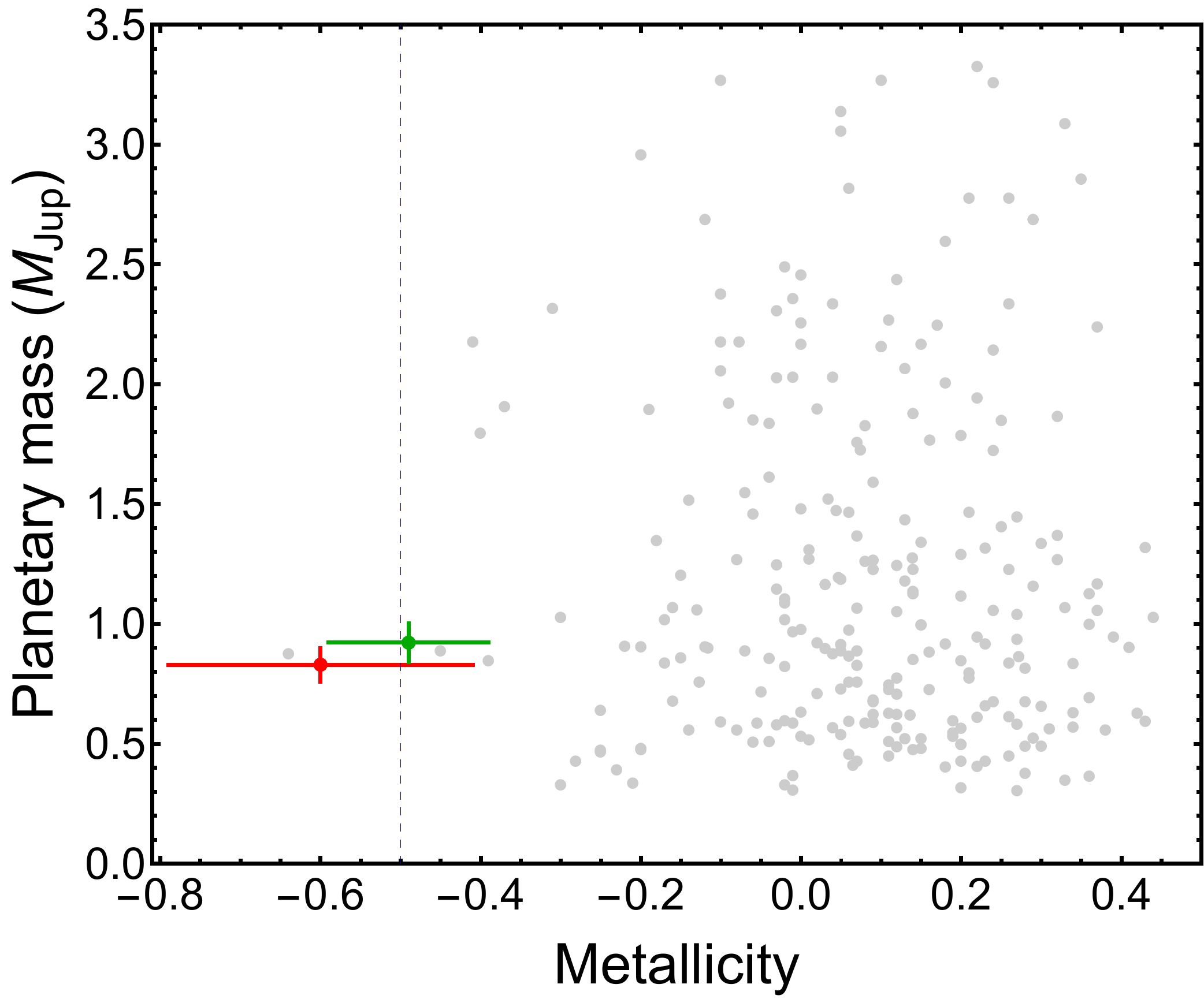}
\caption{Planetary mass versus parent-star metallicity for known transiting planets with $M_{\rm p}>0.3\,M_{\rm Jup}$. The grey points denote values taken from TEPCat. Their error bars have been suppressed for clarity. The WASP-98 planetary system is shown in red \citep{hellier:2014} and green (this work). The dashed vertical line refers to a possible threshold given by the core-accretion theory \citep{mordasini:2012}.}
%Mean density of known transiting exoplanets versus the metallicity of the corresponding parent stars in the range between 0 and $8\,\rho_{\rm Jup}$. The positions of the %planets are represented by circles, whose size is proportional to their radius. Colours indicate the planet equilibrium temperatures. The old and revised positions of WASP-98 %are highlighted. The error bars have been suppressed for clarity. Data taken from TEPCat.}
\label{fig:diagram}
\end{figure}

In this work we have presented new high-quality, broad-band photometric observations of two transit events of the hot Jupiter WASP-98\,b. The observations were performed with the GROND imager, which permits the simultaneous collection of multi-colour light curves. We also obtained three spectra of the parent star with the FEROS spectrograph for confirmation of its spectral characteristics, as it was previously found to be unusually metal-poor for a planet host star \citep{hellier:2014}.
%
%This low metallicity makes the WASP-98 planetary system a possible {\it litmus paper} for the competing theories of giant-planet formation.
%
The new data were analysed, modelled and used to redetermine the orbital ephemeris and physical parameters of the planetary system. Our results, shown in Table\,\ref{tab:derived-params}, are more precise, but still consistent, when compared with those reported by \citet{hellier:2014}. Besides finding that both the star and the planet are slightly larger and more massive than was previously thought, we measured a lower metallicity of the star, \(-0.49 \pm 0.10\)  against \(-0.60 \pm 0.19\), which still makes this planetary system very unusual in the sample of known systems (see Fig.\,\ref{fig:diagram}). 
Considering the strong correlation between the metallicity and giant-planet frequency, if WASP-98\,b was formed via the core-accretion mechanism, it should be a very rare specimen. An alternative formation mechanism, like disk instability, could be advocate for explaining the existence of WASP-98\,b.

%{\bf However, this new value is still not in agreement with the correlation between the metallicity and giant-planet frequency, which is considered a very strong point for the %core-accretion scenario. An alternative formation mechanism, like disk instability, could be advocate for explaining the existence of WASP-98\,b.}
%now closer to the ${\rm [Fe/H]}=-0.5$ threshold for planet formation by core accretion (see discussion in Sect.\,\ref{sec:introduction}) so does not permit a discrimination of %the formation mechanism of this system.

Additionally, we used our new multi-band photometric observations to probe the atmosphere of WASP-98\,b. We found a planetary radius variation between the $r^{\prime}$ and the $z^{\prime}$ bands of $5.5 H$ at a high confidence level of $\sim 6\sigma$. We have compared our observational results with several synthetic spectra obtained with the {\it petitCODE}, considering five possible types of atmosphere. A reasonable fit between the observations and data was only accomplished for the quite unrealistic case in which we suppressed the condensation of Ti and V. Our findings point therefore towards the existence of a mysterious optical absorber that should play an important role in the chemical and physical processes in the atmosphere of WASP-98\,b.

%The transmission spectrum that we have measured is interesting because \citet{evans:2016} have recently found a similar broad-band photometric signature for the much %hotter planet WASP-121\,b, which they attributed to TiO and VO. WASP-121\,b is the first hot Jupiter to have a clear evidence TiO and VO detected in its atmosphere if this %measurement is confirmed by spectroscopy. Previously TiO and VO had not been found for both hotter and cooler planets. If follow-up measurements of WASP-121\,b find %something else, or fail to confirm the presence of TiO and VO this could hint at a similar absorber as we have found evidence for in WASP-98\,b.

%

\section*{Acknowledgements}
This paper is based on observations collected with the MPG 2.2\,m telescope located at the ESO Observatory in La Silla, Chile. Operation of the MPG 2.2\,m telescope is jointly performed by the Max Planck Gesellschaft and the European Southern Observatory. GROND was built by the high-energy group of MPE, in collaboration with the LSW Tautenburg and ESO, and is operated as a PI-instrument at the MPG 2.2\,m telescope. We thank Christoph Mordasini for helpful comments. We also thank Angela Hempel and Wilma Trick for their technical assistance during the observations. JS acknowledges funding from the Leverhulme Trust in the form of a Philip Leverhulme Prize. MG acknowledges the support of the TAsP (Theoretical Astroparticle Physics) Project funded by INFN and would like to thank LM for hospitality during his stay at Max Planck Institute for Astronomy. The reduced light curves presented in this work will be made available at the CDS (http://cdsweb.u-strasbg.fr/).
%We thank the anonymous referee for their useful criticisms and suggestions that helped us to improve the quality of this paper.
The following internet-based resources were used in research for this paper: the ESO Digitized Sky Survey; the NASA Astrophysics Data System; the SIMBAD data base operated at CDS, Strasbourg, France; and the arXiv scientific paper preprint service operated by Cornell University.


%\bibliographystyle{mnras}
%\bibliography{bibliography}

%%%%%%%%%%%%%%%%% APPENDICES %%%%%%%%%%%%%%%%%%%%%

%\appendix

% Don't change these lines
\bsp % typesetting comment
\label{lastpage}
\end{document}